\documentclass{ws-ijqi}

\begin{document}

\catchline{}{}{}{}{}

\title{Systematic study of the PDC speckle structure for quantum imaging applications.   }

\author{ G. BRIDA, M.GENOVESE  \footnote{genovese@inrim.it}, A. MEDA, I. RUO BERCHERA.}

\address{INRIM; strada delle Cacce 91, 10135 Torino, Italy}

\author{E. PREDAZZI}

\address{Dip. Fisica Teorica, Univ. Torino, via P. Giuria 1, 10125 Torino, Italy }

\maketitle

\begin{history}
\received{15-05-2008}
\end{history}

\begin{abstract}
Sub shot noise imaging of weak object by exploiting Parametric Down
Converted light represents a very interesting technological
development. A precise characterization of PDC speckle structure in
dependence of pump beam parameters is a fundamental tool for this
application. In this paper we present a first set of data addressed
to this purpose.

\end{abstract}

\section{Introduction}

Sub Shot Noise (SSN) imaging of weak object by exploiting Parametric
Down Converted light, i.e. to obtain the image of a weak absorbing
object with a level of noise below the minimum threshold that is
unavoidable in the classical framework of light detection,
represents a very interesting technological development
\cite{qi,lug2,dit,and}. A precise characterization of PDC speckle
structure in dependence of pump beam parameters is a fundamental
tool for this application. Indeed, it is fundamental to set the
dimension of the modes coherence areas with respect to the dimension
of the pixels of the ccd camera used for acquiring the images.

Albeit a theory of speckles structure in PDC \cite{lug,mat} has been
developed  and some results in a specific regime have been collected
\cite{dit}, a systematic study of this structure is still missing.
In this paper we present a first set of data addressed to this
purpose.

\section{Theory}\label{theory}

The process of SPDC presents a large bandwidth in the spatial
frequency domain, that  is particularly useful when studying spatial
quantum correlations \cite{MG}. Any pair of transverse modes of the
radiation, characterized by two opposite transverse momenta
$\mathbf{q}$ and $-\mathbf{q}$, are correlated in the photon number,
i.e. they contain, in an ideal situation, the same number of
photons. In the far field zone, the single transverse mode is
characterized by a coherence area, namely the uncertainty on the
emission angle $\vartheta$ ($\tan\vartheta=\lambda q/2\pi$,
$\lambda$ being the wavelength) of the twin photons. It derives from
two effects that participate in the relaxation of the phase matching
condition. On the one side the finite transverse dimension of the
gain area, coinciding with the pump radius $w_{p}$ at low parametric
gain. On the other side the finite longitudinal dimension of the
system, i.e. along the pump propagation direction, that is generally
given by the crystal length $l$. If the first dominates, the
coherence area is related to the Fourier transform of the pump
transverse profile, i.e. $x_{coh}\propto 1/w_{p}$. If the second
dominates, the coherence area is of the order of $(l
\cdot\vartheta)^{-1}$ for small emission angle. The appearance of
the emission is a speckled structure in which the speckles have,
roughly, the dimension of the coherent area and for any speckle at
position $\mathbf{q}$ there exists a symmetrical one in
$-\mathbf{q}$ with equal intensity (see fig. \ref{fig3}).

Omitting some unessential constants, the Hamiltonian describing the
three fields parametric interaction is

\begin{equation}\label{QH_I}
 \widehat{H}_{I}(t)\propto\int_{\mathcal{V}}\chi^{(2)}\;\widehat{\mathbf{E}}^{(+)}_{1}(\mathbf{r},t)\;
 \widehat{\mathbf{E}}^{(+)}_{2}(\mathbf{r},t)\;\widehat{\mathbf{E}}^{(-)}_{p}(\mathbf{r},t)\;d^{3}r
 + h.c
\end{equation}
The pump depletion due to the down-conversion and the absorbtion is
indeed of small entity, unless extremely high intensity laser
sources are used. We shall therefore work in the parametric
approximation, that treats the pump as a classical monochromatic
field propagating linearly along a certain $z$ direction inside the
crystal and having an amplitude transverse profile $A_{p}(\rho)$,
i.e.
\begin{equation}\label{pump}
\mathbf E_{p}\left(\mathbf{r},t\right)\propto A_{p}(\rho)
\;e^{-i(k_{p}z-\omega_{p} t)}\,,
\end{equation}
where $\rho$ is the coordinate vector in the transverse plane to the
propagation direction $z$.

The down-converted fields 1 and 2 are quantized. Their positive-
and negative-frequency part
$\widehat{\mathbf{E}}^{(+)}(\mathbf{r},t)$ and
$\widehat{\mathbf{E}}^{(-)}(\mathbf{r},t)$ is given as an
expansion in plane-wave modes and we find convenient to express
them separating the sum over the wave-vector  into the sum over
its transverse component $\mathbf{q}$ and the frequency $\omega$.
Thus, we have

\begin{equation}\label{dc fields}
\widehat{\mathbf E}^{(+)}_{i}\left(\mathbf{r},t\right)
\propto\sum_{\mathbf{q}_{i},\omega_{i}} e^{i(k_{iz}z-\omega_{i}t)}\;
e^{i\mathbf{q}_{i}\cdot\mathbf{\rho}}\;\widehat{a}_{i}(\mathbf{q}_{i},\omega_{i})
\end{equation}
where $i=1,2$. The third component $k_{iz}$ of the i-th field wave
vector is expressed in terms of the $\mathbf{q}_{i}$ and
$\omega_{i}$ because of the relations
\begin{equation}\label{k_iz}
k_{iz}=\sqrt{k_{i}^{2}-q^{2}}\qquad \mathrm{and} \qquad
k_{i}=\frac{\omega_{i}n_{i}}{c}
\end{equation}
Introducing Eq.s (\ref{pump}) and (\ref{dc fields}) in the
Hamiltonian (\ref{QH_I}) we have

\begin{eqnarray}\label{QH ext}
\widehat{H}_{I}\propto
\sum_{\mathbf{q}_{1},\omega_{1}}\sum_{\mathbf{q}_{2},\omega_{2}}\chi^{(2)}
\int_{0}^{l}e^{i(k_{1z}+k_{2z}-k_{p})z}dz \int_{l_{x}\times
l_{y}}d\mathbf{\rho}\;A_{p}(\mathbf{\rho})\;
e^{i(\mathbf{q}_{1}+\mathbf{q}_{2})\cdot\mathbf{\rho}}\;\nonumber\\e^{i(\omega_{p}-\omega_{1}-\omega_{2})t}
\;\widehat{a}_{1}(\mathbf{q}_{1},\omega_{1})
\widehat{a}_{2}(\mathbf{q}_{2},\omega_{2})+ h.c.
\end{eqnarray}
Here $l$ is the length of the crystal, while $l_{x}\times l_{y}$ is
the area of its transverse surface.

The integral in $dz$ gives a contribution proportional to
$l\cdot\mathrm{sinc}\left[(\Delta k \;l)/2)\right]$ where $\Delta
k\equiv k_{1z}+k_{2z}-k_{p}$. The double integral on the transverse
surface of the crystal gives the Fourier transform of the pump
transverse profile if the crystal is large compared to it. Supposing
a gaussian pump $A_{p}(\rho)=A_{p}e^{-\rho^{2}/w_{p}^{2}}$ the
Hamiltonian (\ref{QH ext}) becomes
\begin{eqnarray}\label{QH ext2}
\widehat{H}_{I}\propto
\sum_{\mathbf{q}_{1},\omega_{1}}\sum_{\mathbf{q}_{2},\omega_{2}}g\cdot
\,\mathrm{sinc}\left[\frac{\Delta
k (\mathbf{q}_{1},\mathbf{q}_{2},\omega_{1},\omega_{2})\cdot l}{2}\right]e^{-(\mathbf{q}_{1}+\mathbf{q}_{2})^{2}\frac{w_{p}^{2}}{4}}\nonumber\\
e^{i(\omega_{p}-\omega_{1}-\omega_{2})t}\,\widehat{a}_{1}(\mathbf{q},\omega_{1})\;\widehat{a}_{2}(-\mathbf{q},\omega_{2})+h.c.
\end{eqnarray}

where we have introduced the dimensionless factor
$g\propto\chi^{(2)}\cdot l\cdot A_{p}$, usually referred to as
parametric gain. Its value determines the number of photons that are
generated in the down conversion process in mode pairs that are
well-phase matched. The evolution of the quantum system guided by
Hamiltonian (\ref{QH ext2}), in the case of relatively high gain
regime, requires a numerical solution and it is discussed in detail
in \cite{lug}. Anyway, in the first order of the perturbation theory
($g\ll1$), the quantum state of the scattered light has the
entangled form

\begin{eqnarray}
\left| \psi \right\rangle &=&\left| \mathrm{vac} \right\rangle+\exp
\left[ -\frac{i}{\hbar }\int \widehat{H}_{I}dt
\right] \left| 0\right\rangle\nonumber\\
&=& \left| \mathrm{vac}
\right\rangle+\sum_{\mathbf{q}_{1},\mathbf{q}_{2}}
\sum_{\omega_{\Omega}}F(\mathbf{q}_{1},\mathbf{q}_{2},\Omega)\left|
1_{\mathbf{q}_{1},\Omega}\right\rangle \left|
1_{\mathbf{q}_{2},-\Omega}\right\rangle,
\end{eqnarray}

\begin{eqnarray}\label{F}
F(\mathbf{q}_{1},\mathbf{q}_{2},\Omega)&=&g\cdot
\,\mathrm{sinc}\left[\frac{\Delta k
(\mathbf{q}_{1},\mathbf{q}_{2},\Omega)\cdot
l}{2}\right]e^{-(\mathbf{q}_{1}+\mathbf{q}_{2})^{2}\frac{w_{p}^{2}}{4}},\nonumber\\
&&\omega_{1}= \omega_{p}/2+\Omega,
\qquad\omega_{2}=\omega_{p}/2-\Omega.
\end{eqnarray}

The coherence area, in the limit of low parametric gain $g$, can be
estimated by the angular structure of the coincidence probability
$\left|F(\mathbf{q}_{1},\mathbf{q}_{2})\right|^{2}$ at some fixed
frequency $\Omega$. As mentioned before, now is clear that we deal
with two functions that enter in the shaping of the coherence area:
the $sinc$ function and the Fourier transformed of the gaussian pump
profile. Since they are multiplied, the narrower determines the
dimension of the area. By expanding linearly the longitudinal wave
detuning around the exact matching point $\Delta
k(\mathbf{q}_{0},\mathbf{q}_{0},\Omega)$ according to relations
(\ref{k_iz}), the $sinc$ function turns out to have a Half Width
Half Maximum of $\Delta q = 2,78/(l\tan\vartheta)$. The HWHM of the
gaussian function, appearing in (\ref{F}), is $\delta q= \sqrt{2\ln(
2)}/w_{p}$. Concerning our experiment, we consider so small emission
angles $\vartheta$ and large enough pump radius $w_{p}$, that we
always work in the region $\delta q/\Delta q< 1$. Therefore, in
principle, the dimension of the coherence area is only determined by
the pump waist.

When moving to higher gain regime, the number of photon pairs
generated in the single mode increases exponentially as $\propto
\sinh^{2}(g)$ i.e. a large number of photons is emitted in the
coherence time along the direction $\vartheta$. In this case, also
the pump amplitude becomes important in the determination of the
speckles dimension. As described in \cite{lug2},  this can be
explained by a qualitative argumentation: inside the crystal, the
cascading effect that causes the exponential growth of the number of
generated photons is enhanced in the region where the pump field
takes its highest value, i.e. close to the center of the beam. Thus,
in high gain regime, most of the photon pairs are produced where the
pump field is closed to its peak value. As a result the effective
region of amplification inside the crystal becomes narrower than the
beam profile. Thus, in the far field one should consider the
speckles as the Fourier transform of the effective gain profile,
that being narrower, produces larger speckles.

A further fundamental consideration for the practical
implementation is that in high gain regime, instead of measuring
the coincidences between two photons by means of two single
photo-detectors, one collects a large portion of the emission by
using for instance a CCD array with a certain fixed exposure time.
Within this time several photons are collected by the single
pixels and the result is an intensity pattern, having the spatial
resolution of the pixel. The coherence area can be evaluated by
the cross-correlation between the signal's and the idler's
intensity patterns. We can define also the auto-correlation
function of the signal intensity pattern itself, since in the
single transverse mode of the signal arm there are many photons.
To be precise the speckle's dimension is better related to the
spread of this function, although the two functions, the cross-
and the auto- correlation, present the same behaviour with respect
to the pump parameters, see \cite{lug}. From the experimental
view-point it is convenient to study the auto-correlation because
of the higher visibility that allows a more accurate estimation of
its size.

\section{Experiment}

\begin{figure}[tbp]
\begin{center}
\includegraphics[angle=0, width=12 cm, height=8 cm]{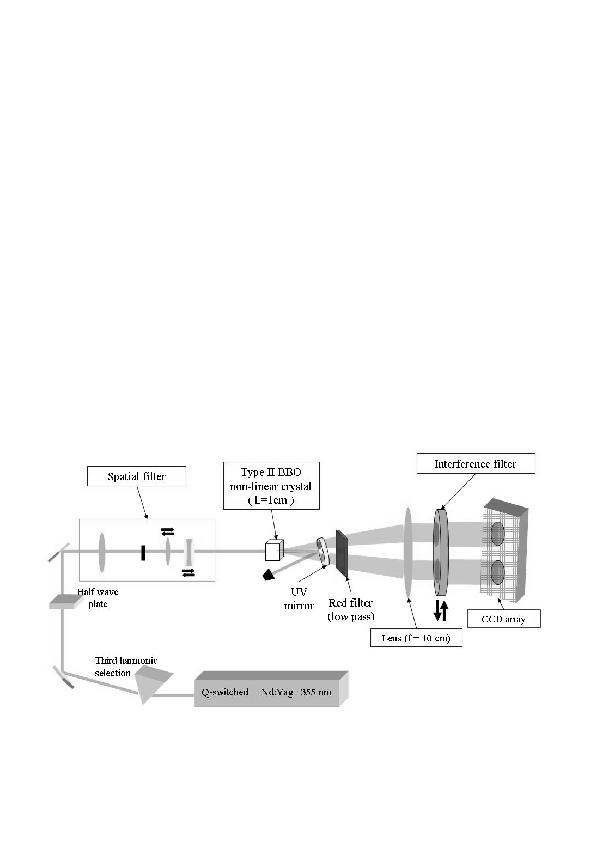}
\caption{Experimental setup. A triplicated Nd-Yag laser beam, after
spatial filtering, produces type II PDC in a BBO crystal, which is
then measured, after an interference filter and pump elimination, by
a CCD camera. }\label{fig2}
\end{center}
\end{figure}

In our setup, Fig.\ref{fig2}, a type II BBO non-linear crystal
($l=1$ cm) is pumped by the third harmonic (wavelength of 355 nm) of
a Q-switched Nd:Yag laser. The pulses have a duration of 5ns with a
repetition rate of 10 Hz and a maximum energy, at the selected
wavelength, of about 200 mJ. The pump beam crosses a spatial filter
(a lens with $f$=50 cm and an iris of 250 $\mu$m of diameter), in
order to eliminate the non-gaussian components  and to collimate it
before the crystal. The diameter of the pump beam entering the
crystal is varied, when necessary, by changing the distance between
two lenses (a biconvex and a biconcave) placed after the spatial
filter. After the crystal, the pump is stopped by a UV mirror,
transparent to the visible, and by a low frequency-pass filter. The
down converted photons (signal and idler) pass through a lens of 5
cm of diameter ($f=10$ cm) and an interference filter centered at
the degeneracy $\lambda$=710 nm (10nm bandwidth) and finally
measured by a CCD camera. We used a $1340X400$ CCD array, Princeton
Pixis:400BR (pixel size of 20 $\mu$m), with high quantum efficiency
(80\%) and low noise (5 electrons/pixel). The far field is observed
at the focal plane of the lens in a $f-f$ optical configuration,
that ensures that we image the Fourier transform of the crystal exit
surface. Therefore a single transverse wavevector $\mathbf{q}$ is
associated to a single point $\mathbf{x}=(\lambda f/2\pi)\mathbf{q}$
in the detection plane. The CCD acquisition time is set to 90 ms, so
that each frame corresponds to the PDC generated by a single shot of
the laser.

\begin{figure}[tbp]
\begin{center}
\includegraphics[angle=0, width=15 cm, height=10 cm]{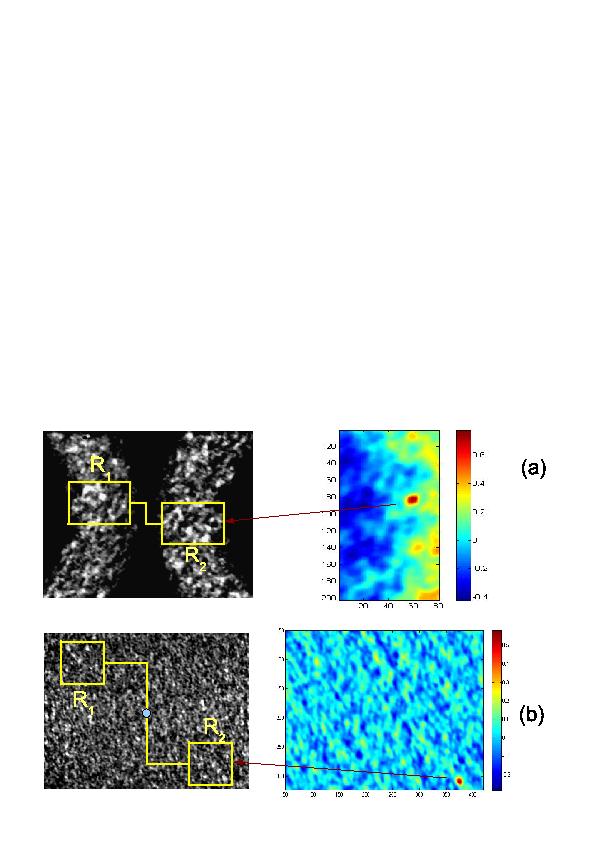}
\caption{Spatial correlation between signal and idler down
converted light with interference filter around degeneracy(a) and
without a narrow filter (b). In each raw the first inset is a ccd
image, the second a plot of the cross correlation function. The
axis report the pixels position in the CCD array.
$C_{12}(\mathbf{\xi})$. }\label{fig3}
\end{center}
\end{figure}

Looking at the images, we can appreciate the speckled structure and
a certain level of correlation of the speckles intensity between the
signal and idler arms (Fig. \ref{fig3}). Let us define
$N_{R}(\mathbf{x})$ the intensity level, proportional to the number
of photons, registered by the pixel in the position $\mathbf{x}$ of
the region $R$. $\delta N_{R}(\mathbf{x})= N_{R}(\mathbf{x})-\langle
N_{R}(\mathbf{x})\rangle$ is the fluctuation around the mean value
that is estimated as $\langle
N_{R}(\mathbf{x})\rangle=(1/n)\sum_{\mathbf{x}} N_{R}(\mathbf{x})$,
with $n$ the number of pixels. We evaluate the normalized spatial
cross-correlations of the intensity fluctuations in an arbitrary
region $R_{1}$, belonging to the signal portion of the image, and in
the symmetric region $R_{2}$ (see Fig. \ref{fig3}) belonging to the
idler portion:

\begin{equation}\label{cross-corr}
 C_{12}(\mathbf{\xi})=\frac{\left\langle\delta
 N_{R_{1}}(\mathbf{x})\delta N_{R_{2}}(\mathbf{-x+\xi})\right\rangle}
 {\sqrt{\left\langle \delta N_{R_{1}}(\mathbf{x})^{ 2}\right\rangle\left\langle\delta N_{R_{2}}(\mathbf{-x+\xi})^{2}\right\rangle}}
\end{equation}

where  $\xi$ is the displacement vector that assumes discrete
values. $C_{12}$ reaches a peak of about 0,9 in Fig. \ref{fig3}-a,
that indicates a good level of spatial correlation between signal
and idler. In figure \ref{fig3}-b the value is around 0,6 because
here we did not put the interference filter in front of the CCD
camera, allowing more background light to enter, and the signal and
idler component are not separated. It is worth to emphasize that
$C_{12}(\mathbf{\xi})$ is not a index of the correlation at the
quantum level and it can not be used to discriminate the SSN
condition. We are interested mainly in the width of the peak, that
indicate the coherence area. On the other hand, the SSN condition is
checked by a measurement of $ \sigma^2  = < (N_{R_{1}}
-N_{R_{2}})^2>-<N_{R_{1}}-N_{R_{2}}>^2 < 1$: our preliminary data
show that this result is reached for some images without background
subtraction; a detailed discussion of this investigation is
postponed to a forthcoming paper.

Since the experimental cross-correlation is characterized by a
non-optimal visibility, due to the losses in optical paths and
noise, we preferred to estimate the speckle's size, by evaluating
the auto-correlation of a single region R of the signal intensity
pattern.

\begin{equation}\label{auto-corr}
 C(\mathbf{\xi})=\frac{\left\langle\delta
 N_{R}(\mathbf{x})\delta N_{R}(\mathbf{x+\xi})\right\rangle}
 {\sqrt{\left\langle \delta N_{R}(\mathbf{x})^{ 2}\right\rangle\left\langle\delta
 N_{R}(\mathbf{x+\xi})^{2}\right\rangle}}.
\end{equation}

\begin{figure}[tbp]
\begin{center}
\includegraphics[angle=0, width=15 cm, height=12 cm]{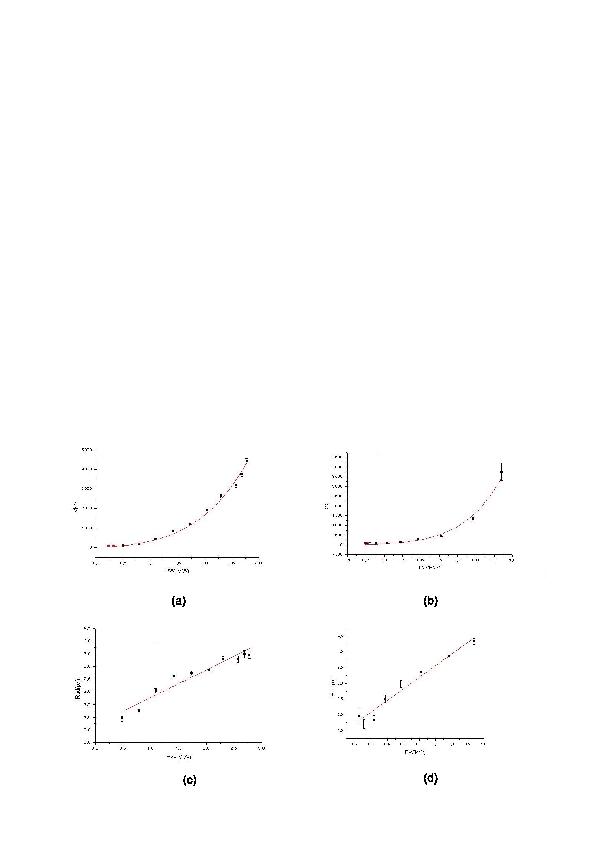}
\caption{The top of the figure represents the mean number of
photon/pixel collected by the CCD in one shot of the laser (5ns),
function of the pump pulse power in MW. The bottom represents the
dependence of the speckle radius (in pixel) from the pump power. In
(a) and (c) the pump diameter is 1,3mm, while in (b) and (d) is
0.95mm. }\label{fig4}
\end{center}
\end{figure}

First off all we investigate the gain region in which we are
working, in order to ensure the possibility to reach a sufficient
non-linear cascading effect in the photon pairs production inside
the crystal. Fig. \ref{fig4} shows the mean photon number $\langle
N_{R}(\mathbf{x})\rangle$ as function of the pump power $PW$. Any
point is averaged over several tenths of frames in order to reduce
the uncertainty. Our laser presents in fact 20\% fluctuations of the
power from pulse-to-pulse. Since the mean number of photons is
proportional to $ \sinh^{2}(g)$, the fluctuations of the pump
generate large fluctuations in the photons number from
frame-to-frame. The diameter of the pump is fixed and the power is
varied by the delay between the Q-switch turn-on and the lamp flash.
The calibration curve delay-power has been measured by a power meter
and we observed a reproducibility with uncertainty around 10\%.

In Fig. \ref{fig4}-a the pump diameter is around 1,3mm and in Fig.
\ref{fig4}-b is 0,95mm. Although the pump power used for the case
(b) is smaller than case (a), the intensity results higher in the
case (b) because of the reduced radial dimension. It must be noticed
that we are constrained in the range of intensities of the pump. For
high intensities we are limited by the damage threshold of optical
components, for low intensities by the visibility of the speckle
structure because we collect a lot of temporal modes in the same
frames.

The data are fitted by the equation $\langle
N_{R}\rangle=k\cdot\sinh^{2}(\sigma \sqrt{p})$. k depends from the
number of modes while $\sigma\sqrt{P}=g$. The experimental values,
mediated on three different acquisitions, are $k=31.48$
 and
$\sigma=1.91$ for the case (a). In the case (b) we obtain $k=1,10$
and $\sigma=4,87$; the higher value of $\sigma$ is due to the
increased intensity. Therefore, in the case (a) we have a gain $g$
from 1,5 to 3,5 and in the case (b) from 1,9 to 5. Thus, we are in
a non-linear regime where we should expect a dependence of the
speckles size from the amplitude of the pump.

Fig. \ref{fig4}-c and \ref{fig4}-d show the trend of the radius of
the coherence area with the pump power, again at the two different
diameters of the pump. Actually we observe an increasing of the
radius, that is predicted by the considerations exposed in section
\ref{theory}. We consider, just as indicative, a linear fit
$y=a(x-b)$, obtaining $a=1,25$ and $b=-0,63$ for (c) and $a=3.7$ and
$b=-0,27$ for (d). The higher slope of (c) is qualitatively
explained by the higher gain.

\begin{figure}[tbp]
\begin{center}
\includegraphics[angle=0, width=15 cm, height=10 cm]{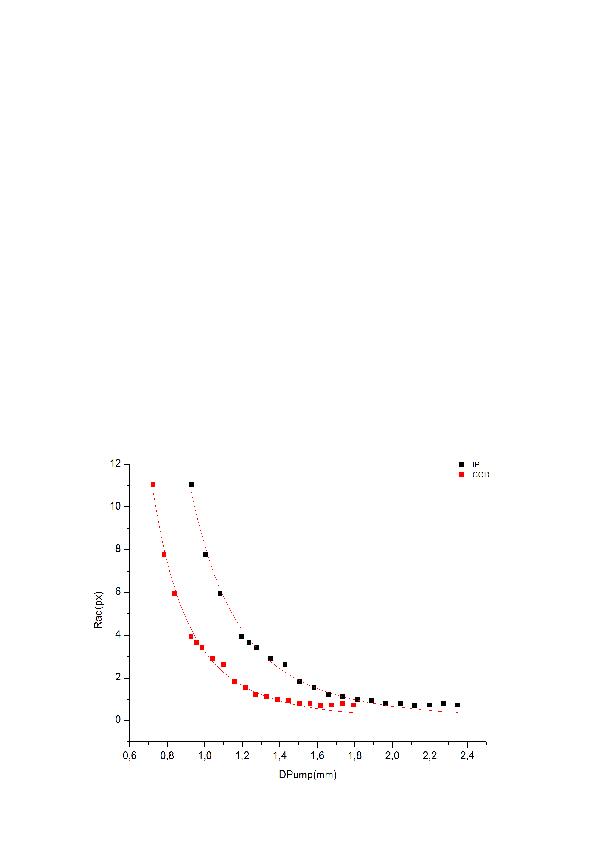}
\caption{Observed dependence of the radius (in pixels) of the
speckles from the pump diameter. The two curves differs in the
estimation of the pump diameter: in one case it was measured with an
impact paper (IP) and in the other case with a CCD. }\label{fig5}
\end{center}
\end{figure}

Finally, we investigate the dependence of the radius of the
speckles from the pump diameter, as shown in figure \ref{fig5}. We
fix the power of the laser to 0,78 MW and we change the diameter
varying the distance between the two collimating lenses (see
figure \ref{fig2}). The two curves differ just in the estimation
of the pump diameter: in one case we measured it with an impact
paper (IP) and in the other case with a CCD. We perform a fit in
the form $y=a\cdot x^{b}$ obtaining $a=(8,1\pm0,1)$ and
$b=(-3,61\pm0,09)$ for the IP curve and $a=(3,22\pm0,07)$ and
$b=(-3,73 \pm 0,09)$ for the CCD curve. Despite the different way
of pump size estimation, reflected in the value of the parameter
a, the coefficients b are compatible. The theory, in low gain
regime, provides that the radius of the speckles is proportional
to the inverse of the pump size ($w_{p}$), i.e. $b=-1$. The
estimated value of b, in our case, confirms the role of the high
gain regime in the speckle size. In fact, together with the
reduction of the pump diameter, the gain increases, and thus the
effective gain area is more reduced again. This effect impresses
upon the speckles size a stronger dependence with respect to the
pump size.

\section{Conclusion}

In this paper we have presented a first experimental study of the
size of the coherence area in PDC in the high gain regime. We have
shown that the speckles present, not only a dependence on the pump
diameter, as in the usual low gain regime, but also a strong
dependence from the pump intensity. This result, i.e. the
understanding of the behaviour of the coherence area in the high
gain regime, is fundamental for the innovative application in the
field of quantum imaging \cite{qi},  but also in general for quantum
metrology, quantum information, etc. A comparison with these data
will represent a challenge for the theoretical models describing
this phenomenon.

\subsection{ Acknowledgments}
This work has been supported by MIUR (PRIN 2005023443-002), by
07-02-91581-ASP, Compagnia di San Paolo Foundation,  EU project
QuCandela and by Regione Piemonte (E14). Thanks are due to
Alessandra Gatti, Ottavia Jedrkiewicz and Luigi Lugiato for useful
discussions.

\markboth{Taylor \& Francis and I.T. Consultant}{Journal of Modern Optics}

\label{lastpage}

\end{document}